\pdfoutput=1 
 \documentclass[nohyper,12pt,letterpaper]{JHEP3}
 \usepackage{epsfig}
 
 \title{String Embeddings of the Pentagon}

 \author{S.\,Echols\\
 Department of Physics \\
 Cal Poly State University, San Luis Obispo, CA 93407\\
 E-mail: \email{sechols@calpoly.edu}}

\abstract{The Pentagon Model is an explicit supersymmetric extension of the
Standard Model, which involves a new strongly-interacting $SU(5)$
gauge theory at TeV-scale energies.  We discuss embeddings of the Pentagon Model into string theory, specifically
$\mathcal{N} =1$ supersymmetric type IIa intersecting D-brane
models, M-theory compactifications of $G_2$ holonomy, and heterotic
orbifold constructions.}

\begin{document}

 \section{\bf Introduction}

The string model-building program has a number of goals.  First, if
completely realistic models are found, this would provide a proof
that string theory may be a unified theory of all particles and
interactions.  Further, the study of the surviving low-energy
spectra of various string models might lead to the identification of
general patterns (such as symmetries or exotic particle content)
present in a large class of realistic vacua.  Additionally, it might
lead to new ideas for addressing problems such as CP violation,
fermion mass mixings, or even dark matter and dark energy.  Perhaps
most significant is the hope that the discipline will lead to
experimentally testable predictions.  The last of these is
especially provocative at a time when we find ourselves on the
verge of a plethora of new data from the LHC.

The Pentagon Model of TeV physics successfully addresses a number of low energy
phenomenological issues, and we would therefore like to find it as an effective
field theory of a string construction.  Such a search is the subject
of this paper.

Though previous search \cite{pentagonGUT} has produced promising results towards embedding the Pentagon into a grand unified theory (GUT),  the method nevertheless relies on arguments involving operators at
the Planck scale where in reality the theory breaks down and becomes
unreliable.  Thus, the purpose of this paper is to explore the
possibility of embedding the Pentagon Model into a string theory
directly.  In practice this translates into choosing one specific
string theory with a given geometry and turning the crank to find
the resulting particle spectrum of that theory, and comparing it
with the Pentagon.  There are currently five different types of
string theories for which it is known how to calculate the
low-energy chiral spectrum: orbifold constructions in heterotic
string theory, $G_2$ compactifications of M-theory, intersecting
D-brane models in either type IIA or IIB string theory, and F-theory
models.  In this chapter we will consider the first three of these
approaches.  Type IIB and their dual F-theory models might certainly
be of interest, but are left to future work.

Though a search for the Pentagon has never before been performed, it
should be noted that each of these approaches has yielded only
mediocre results in previous searches for the standard model and
various GUTs.  While many models have been discovered which may
contain the desired particle content and gauge symmetries, one must
also contend with other issues such as problems with the existence
of chiral exotics, symmetry breaking and Higgs fields, and finding
proper $U(1)$ charges and Yukawa couplings.  While extensive
research has been devoted to these questions, only perhaps a handful
of models have satisfactorily addressed all of these issues.  As the
purpose of this search is merely to establish the viability of the
existence of the Pentagon, our strategy has been to search for
models that are `at least as good as state of the art'.  In other
words, we must begin by searching the various string theories for
our desired particle content.  If we were to find a massless
particle spectrum corresponding to that of the Pentagon, we would
then turn our attention to the phenomenological aspects of these
models.

Unfortunately, we have found that the existence of the Pentagon as a
low energy spectrum of these theories is impossible at worst and
inconclusive at best.  The difficulty seems to arise due to the
requirement that we obtain both chiral and vector-like particles, as
will be discussed.  The paper is constructed as follows: We first briefly review the contents of the Pentagon Model in section 2.  In
section 3 we will consider models of $\mathcal{N}=1$ globally
supersymmetric type IIA intersecting D6-brane construction. In
\cite{CPS}, Cveti\v{c}, Papadimitriou, and Shiu (CPS) have performed
an extensive search for $\mathcal{N}=1$ supersymmetric three-family
$SU(5)$ Grand Unified Models in the type IIA context, and we
therefore use their work as the starting point for our search.  In
Section 4 we consider the $G_2$ lift of these models onto eleven
dimensional M-theory compactifications.  We find a possible
candidate for the Pentagon in this context, but cannot provide a
proof for its existence.  Section 5 is devoted to heterotic orbifold
models.  It appears difficult to find a consistent model supporting
an $SU(5) \times SU(5)$ gauge group with charged matter in the
bifundamental representation, so we instead focus our search for the
chiral spectrum of the Pyramid Model.  The results of some of the
more promising models are listed but we were unable to find an exact
replication of the low energy model, though we were unable to rule
out the possibility.  In Section 6 we write some concluding remarks.

\section{The Pentagon}

\subsection{The Original Model}

The Pentagon is a supersymmetric model of TeV scale physics
\cite{pentagon,remodel}, whose foundation is the Minimally
Supersymmetric Standard Model (MSSM).  The Pentagon model was origianally constructed to address standard issues with the MSSM, such as SUSY breaking, the $\mu$ problem, the flavor problem, CP violation, and baryon violation .  In
addition to an $SU(5)$ grand unified version of the MSSM, a new
strongly interacting `Pentagon' $SU(5)$ super-QCD with five flavors
of pentaquarks is introduced as a hidden sector which mediates SUSY
breaking through Standard Model gauge couplings. An
hypothetical meta-stable $N_F=N_C=5$ vacuum of the theory is used
to employ the SUSY breaking mechanism of Intriligator, Seiberg, and
Shih (ISS) to construct an effective theory for Cosmological SUSY
breaking (CSB). It naturally introduces a $\mu$ term of the right
order of magnitude, contains a discrete R-symmetry which eliminates
all unwanted dimension 4 and 5 Baryon and Lepton violating
operators, and resolves the SUSY flavor problem.  Strong CP violating
phases remain in the model (in addition to standard neutrino see-saw and CKM matrix phases), but these are potentially addressed with
the addition of an axion.

The Lagrangian
 of the Pentagon model contains several pieces.  The standard MSSM Lagrangian is
implemented as usual: the kinetic energy terms for the matter and
Higgs fields arise in the Kahler potential,
$$
\mathcal{L}_1=d^4 \theta [P^*e^VP+Q^*e^VQ+L^*e^VL+(\bar{U})^*e^V \bar{U}+(\bar{D})^*e^V \bar{D}+(\bar{E})^*e^V \bar{E}
$$
and the gauge superpotential produces the kinetic terms for the gauge fields and gauginos,
$$
\mathcal{L}_2=\int d^2\theta (\sum \tau_i W^i_{\alpha})^2.
$$
Yukawa couplings for the Standard Model fermions and a mass term for
the Higgsino are contained in the superpotential,
$$
\mathcal{L}_3=\int d^2\theta \lambda_u H_u Q \bar{U} + \lambda_d H_d Q \bar{D} + \lambda_L H_d L \bar{E} + {\lambda^{mn} \over M_U} L_m L_n H_u^2 +h.c.
$$
In addition to the MSSM Lagrangian, the Pentagon model includes an
additional superpotential for the pentaquarks (transforming as $P
\sim [5,\bar{5}]$ and $\tilde{P} \sim [\bar{5},5]$ under the
$SU(5)_P \times SU(5)_{GUT}$ gauge group) and an additional singlet
field $S$ with discrete R-charge 2:
$$
\mathcal{L}_4=\int d^2\theta P_i^A \tilde{P}^j_A(m_{ISS}\delta^i_j + g_S S Y_j^i) + g_{\mu} S H_u H_d + g_T S^3.
$$
The scale $M_U$ is taken to lie in the range $M_U \sim  10^{14} -
10^{15}$ GeV to successfully implement the neutrino seesaw effect.
$m_{ISS}$ is assumed to be induced by CSB in the UV sector of the
theory (we will discuss CSB further in the next section),
$$
m_{ISS}=\gamma { \lambda^ {1/4} M_P \over \Lambda_5}.
$$
$\lambda$ is the cosmological constant, $M_P$ the Planck mass, and
$\Lambda_5$ the confinement scale of the Pentagon gauge group.  To
be consistent with CSB, $\Lambda_5 \sim 1.5$ TeV. $\gamma$ is an
unknown constant of order one.  ISS proved that for a theory of SUSY
QCD with $N_C+1 \leq N_F \leq {3 N_C \over 2}$, the mass term
$m_{ISS} {\rm Tr} P \tilde{P}$ induces a meta-stable SUSY violating
ground state with SUSY order parameter $F \sim m_{ISS}
\Lambda_{N_C}$\footnote{The analysis is only under analytic control if $m_{ISS} << \Lambda_{N_c}$} \cite{iss}.  They further argued that a similar
meta-stable state might exist for a theory with $N_F=N_C$, though its
properties could not be calculated analytically.  The Pentagon
therefore has a stationary point of its effective potential with a
non-zero vacuum energy of order $m_{ISS}^2 \Lambda_5^2$.  SUSY
breaking is communicated to the Standard Model via two mechanisms.
The dominant contribution to gaugino masses as well as the masses of
the squarks and sleptons is through conventional gauge mediation. The
Higgs superfields also contribute tree level masses to squarks and
sleptons due to non-zero F terms.

The singlet field $S$ is thought to be the remnant of an $SU(5)$
adjoint, transforming like the hypercharge generator of $SU(3)
\times SU(2) \times U(1)$.  Its coupling to the Standard Model
therefore implies that the GUT $SU(5)$ is broken to $SU(3) \times
SU(2) \times U(1)$.  It also ties the properties of the meta-stable
SUSY violating vacuum to electroweak symmetry breaking through its
F-term, predicting $SU(2) \times U(1) \rightarrow U(1)_{EM}$ with
$|h_u| \sim |h_d| \sim \Lambda_5$, $ \tan \beta \sim 1$.
Furthermore, the VEV of $S$ can give rise to a natural $\mu$ term.

The SUSic $m_{ISS} \rightarrow 0$ limit of the theory admits an anomaly free R-symmetry which is identified with the discrete $Z_N$ R-symmetry
required by the rules of CSB. The SUSY degrees of freedom
transform non-trivially under an R-symmetry, it follows that $N=4$
to accommodate all terms in the superpotential. In models of CSB,
the discrete R-symmetry guarantees Poincar\'{e} invariance; it also
has the effect of preventing all unwanted dimension 4 and 5 baryon
and lepton violating operators leading to proton decay\footnote{This symmetry is explicitly broken by the ISS mass term, but the arguments of CSB lead us to believe that these operators will still be supressed when the cosmological constant is non-zero, i.e. in the SUSY broken theory.}.  The $Z_4$
also forbids various dimension 5 flavor combinations, so quark and
lepton flavor changing processes arise from dimension 6 operators.
Thus, similar to generic gauge mediated models, flavor changing
neutral currents are suppressed below experimental limits.

R-parity preservation implies that the LSP is the gravitino.
Estimates of the scale of SUSY breaking give a gravitino mass of
order $5 \times 10^{-3}$ eV, consistent with Big Bang
Nucleosynthesis.  It is far too small, however, to be a viable dark
matter candidate, and it is strongly coupled enough that the NLSP
will decay too rapidly to be of cosmological importance.  Thus there
is no conventional MSSM dark matter candidate.  On the other hand,
ISS show that the Pentabaryons, dimension one fields made of five
pentaquarks, have a non-vanishing expectation value in the
meta-stable vacuum.  Pentabaryon number is therefore spontaneously
broken,
$$
\langle B \rangle =  \Lambda_5 e^{ib/\Lambda_5},
$$
and the associated Goldstone boson, the penton, is cosmologically
long-lived. If the penta-baryon asymmetry produced in the early universe
is sufficiently large, the penton can be the dark matter.
Furthermore, the pentabaryon and baryon numbers are coupled by QCD
interactions, providing a possible connection between the dark
matter and the observed baryon asymmetry of the universe.  We will
discuss the issues of dark matter and baryogenesis further in the
next section.

\subsection{The Pyramid}

After its invention, it was noticed that the Pentagon model may
suffer from a number of troubling issues.  Most importantly,
$\Lambda_5 \sim 1.5$ leads to a Landau pole before gauge coupling
unification.  In fact, a calculation of the two loop $\beta$
functions for the running of Standard Model couplings requires both
$\Lambda_5, m_{ISS} > 10^3$ TeV \cite{JeffJones}. This is
inconsistent with the conditions of CSB.  Another problem has to do
with stellar phenomenology.  The penton gains mass through a
dimension 7 operator; if the scale associated with this operator is
too large, stars will produce an overabundance of pentons leading to
unobserved stellar cooling \cite{BanksHaber}.

The successor of the Pentagon, the Pyramid model, was constructed to
address these issues \cite{pyramid}.  The Pyramid model employs an
$SU(3)^4$ gauge symmetry, each factor being represented by the
vertices of a pyramid quiver diagram.  Standard Model particles
exist as broken multiplets running around the base of the
pyramid--singlets of a new Pyramid $SU(3)_P$ gauge group, but
fitting into complete multiplets of a conventional trinification
GUT.  In such models, a single generation of fermions comes in the
representation
$$
(3,1,\bar{3}) \oplus (\bar{3},3,1) \oplus (1,\bar{3},3)
$$
under the trinification $SU_1(3) \times SU_2(3) \times SU_3(3)$.
This respects a $Z_3$ permutation symmetry, and can be embedded
precisely into the {\bf 27} of $E_6$.  $SU_3(3)$ is identified with the
color symmetry of the Standard Model, electroweak symmetry comes
from an $SU(2)$ subgroup of $SU_2(3)$, and hypercharge is a linear
combination of generators from both $SU_2(3)$ and $SU_1(3)$.  Gauge
coupling unification is guaranteed if all matter comes in complete
representations of $SU(3)^3 \times Z_3$ and this symmetry is
preserved by Yukawa couplings.

Analogous to pentaquarks, trianons are introduced to implement the
ISS mechanism of meta-stable SUSY breaking and to mediate SUSY
breaking to the MSSM.  Trianons transform under both the Pyramid
$SU(3)_P$ and the trinification symmetry:
$$
\mathcal{T}_1 + \bar{\mathcal{T}}_1 = (3,1,1;\bar{3}) + (\bar{3},1,1;3)
$$
$$
\mathcal{T}_2 + \bar{\mathcal{T}}_2 = (1,3,1;\bar{3}) + (1,\bar{3},1;3)
$$
$$
\mathcal{T}_3 + \bar{\mathcal{T}}_3 = (1,1,3;\bar{3}) + (1,1,\bar{3};3)
$$
Because they respect the $Z_3$ symmetry, one loop perturbative
coupling unification is preserved.

The remainder of the construction of the theory is in complete
parallel with the Pentagon model.  The singlet field $S$ can give rise to a $\mu$ term, and its F-terms gives a VEV to the meson fields that are responsible for electroweak symmetry breaking.  A discrete R-symmetry exists as a consequence of CSB which forbids all dangerous dimension 4 and 5 operators.  Gaugino and squark masses are estimated to lie in an acceptable range for
phenomenology.  The pyrmabaryons themselves are expected to be the
prime dark matter candidate, although spontaneous breaking of
pyrmabaryon number does occur in the model.  The Goldstones of this
broken symmetry are called the pyrmions which, in contrast to the
pentons, avoid constraints from stellar cooling.

Although the majority of this paper is devoted toward developing the
Pentagon model, we do address how the Pyramid model can be extended
to accommodate these developments.

\section{Intersecting type IIA D-branes}

\subsection{Brief review}

Intersecting D-brane models provide a very nice geometric picture
for some of the fundamental ingredients of any low energy effective
field threory\footnote{For a review, see \cite{DbraneReview}.}.  In
particular, they provide a mechanism for generating not only gauge
symmetries but also chiral fermions, where family replication is
achieved by multiple topological intersection numbers of various
D-branes.  To be more specific, the spectrum of open strings
stretched between the intersecting D-branes contains the chiral
particles which are localized at the intersections.  In this section
we will consider specifically the construction of four dimensional
$\mathcal{N}=1$ supersymmetric type IIA orientifolds with D6-branes
intersecting at angles.

Type IIA superstring theory exists in 10 space-time dimensions, six
of which must be compactified to make contact with the observed
world.  The theory contains both closed and open strings as well as
extended charged objects of higher dimension--the D-branes.
Fluctuations of these objects can be described as open strings
attached to the D-branes. The endpoints of the strings give
Chan-Paton factors, which can be viewed as a $U(1)$ gauge field with
momentum only along (and therefore confined to) the brane.  By
placing $N$ D-branes on top of each other the gauge fields on the
branes will transform in the adjoint representation of the gauge
group $U(N)$.  If these fields are carried by D6-branes, three
dimensions must remain uncompactified for these fields to be free to
move in four dimensional Minkowski space-time.  This means that in
the six dimensional transverse compact space the branes are three
dimensional and wrap a three dimensional cycle.  In general, two
such branes will intersect at a point in the compactified space.

An open string extended between the two branes can be shown to have
only one fermionic degree of freedom.  Taking into account an open
string with the opposite orientation between the two D6 branes, one
is left with two fermionic degrees of freedom corresponding to one
chiral Weyl fermion from the four dimensional point of view.  In the
same way, strings extended at the intersection of two stacks of
branes, with $N$ and $M$ D6-branes per stack respectively, will give
rise to a chiral fermion transforming in the bifundamental
representation of $U(N) \times U(M)$.  While the gauge fields are
confined to the branes, gravity still propagates throughout the
bulk.  Thus, the D-branes interact gravitationally, which means that
they will contribute positively to the vacuum energy.  To cancel
this contribution, we must introduce negative tension objects known
as orientifold planes.  Both the D-branes and the orientifold planes
carry R-R charge, which must vanish for consistency.  This gives
rise to tadpole cancellation conditions, which must be satisfied
along with certain supersymmetry conditions for the theory to be
consistent.

The simplest compactification scheme is six dimensional toroidal
compactification factorized as the product of three rectangular
two-tori, $T^6=T^2 \times T^2 \times T^2$, and to assume that the
D6-branes are the products of one-cycles in each of the three
two-tori.  This allows us to specify the branes by wrapping numbers
$(n^i,m^i)$ along the fundamental cycles $[a^i]$ and $[b^i]$ on the
$i$th $T^2$.  Next we introduce the orientifold O6-plane, and allow
it to wrap along each of the $[a^i]$ cycles (as well as the
transverse uncompactified space).  The introduction of the
orientifold plane mods the theory by world-sheet parity as well as
an anti-holomorphic involution, so that the 06-plane is localized at
the fixed plane of the local reflection $(n^i,m^i) \rightarrow
(n^i,-m^i)$.  However, in this scenario, if the D6-branes do not lie
entirely parallel to the 06-plane everywhere, the tension of these
branes in the perpendicular directions cannot be canceled.  Thus, no
non-trivial globally supersymmetric consistent models can be
constructed on these manifolds.

This problem can be alleviated by extending the orientifold planes
into all perpendicular directions via
orbifolding\cite{Cvetic:2001nr}.  The simplest examples of such
models are orientifolds of toroidal type IIA orbifolds $T^2 \times
T^2 \times T^2 /(Z_2 \times Z_2)$.  Using the notation of
\cite{Cvetic:2001nr,CPS}, the orbifold twists are $v=(1/2, -1/2, 0)$
and $w=(0,1/2,-1/2)$, acting on the complex coordinates of the three
two-tori as
$$
\Theta: (z_1,z_2,z_3) \rightarrow (-z_1,-z_2,z_3)
$$
$$
\omega: (z_1,z_2,z_3) \rightarrow (z_1,-z_2,-z_3).
$$
Orientifolding mods the theory by the orientifold action $\Omega R$,
where $\Omega$ is world-sheet parity and $R$ acts as
$$
R: (z_1,z_2,z_3) \rightarrow (\bar{z_1},\bar{z_2},\bar{z_3}).
$$
As with the case of toroidal compactification, the action of the
orientifold requires the O6-plane to lie along the three $[a_i]$
cycles.  However, orbifolding creates three new classes of
O6-planes, each associated with the combined action of the
orientifold and the orbifold:
$$
\Omega R: [\Pi_1]=[a_1][a_2][a_3], \hspace{.2in}
\Omega R \Theta:   [\Pi_2]=[b_1][b_2][a_3],
$$
$$
\Omega R \omega: [\Pi_3]= [a_1][b_2][b_3], \hspace{.2in}
\Omega R \Theta \omega: [\Pi_4]= [b_1][a_2][b_3].
$$

The complex structure of the tori is arbitrary but must be
consistent with the orientifold projection.  This admits only two
choices, each torus may be rectangular (with the lattice vectors
$e_1 \perp e_2$) or tilted such that $e_1'=e_1+e_2/2,e_2'=e_2$.  To
describe both choices in a common notation, a generic one cycle can
be written as $n_a^i[a_i]+l_a^i[b_i]$, with $l_a^i=m_a^i$ for a
rectangular torus and $l_a^i=2m_a^i+n_a^i$ for a tilted torus.  The
homology class of a three cycle is just the product of three one
cycles,
$$
[\Pi_a]=\prod^3_{i=1} (n_a^i[a_i]+2^{-\beta^i}l_a^i[b_i])
$$
where the factor $2^{-\beta^i}$ is included to account for tilted
tori ($\beta_i =1$ if the $i$th torus is tilted, zero otherwise).
The orientifold action maps a one cycle ($n_a^i,l_a^i$) to its image
($n_a^i,-l_a^i$), thus for any stack of D-branes we must also
include its image
$$
[\Pi_a']=\prod^3_{i=1} (n_a^i[a_i]-2^{-\beta^i}l_a^i[b_i]).
$$
Finally, we define
$$
[\Pi_{O6}]=8[\Pi_1]-2^{3-\beta_1-\beta_2}[\Pi_2]-2^{3-\beta_2-\beta_3}[\Pi_3]-2^{3-\beta_1-\beta_3}[\Pi_4].
$$
The coefficients reflect the number of images of each O6 plane that
must be included.

With these definitions we are equipped to consider the open-string
spectrum of the theory. Chiral sectors are defined by the objects
between which the strings in the sector are extended. Adjoint fields
are given by strings with endpoints on a single brane, thus the
gauge group is found in the $aa$ sector.  As mentioned, in toroidal
theory a stack of $N_a$ D6-branes gives rise to a $U(N_a)$ gauge
group.  In the orbifold theory, the $\Theta$ action breaks this to
$U(N_a/2) \times U(N_a/2)$, and the $\omega$ action identifies these
factors, leaving the gauge group $U(N_a/2)$. However, in the special
case of branes coincidental with some of the O6-planes, the symmetry
is enhanced to a $USp(N_a)$ gauge group. Massless strings extended
between these branes will necessarily be vector-like, and so they
have gained the name `filler branes' because they can contribute an
RR tadpole charge without adding to the particle spectrum.

The $ab+ba$ sector gives chiral supermultiplets in the
bi-fundamental representation $(N_a/2,\overline{N_b/2})$.  The
multiplicity of these states is given by the topological
intersection number
$$
I_{ab}=[\Pi_a][\Pi_b]=2^{-k}\prod^3_{i=1}(n_a^il_b^i-n_b^il_a^i)
$$
with $k=\beta_1+\beta_2+\beta_3$.  Similarly, the $ab'+b'a$ sector
(the prime indicates the $\Omega R$ image) gives $I_{ab'}$ chiral
fields in the representation $(N_a/2,N_b/2)$, with
$$
I_{ab'}=[\Pi_a][\Pi_b']=-2^{-k}\prod^3_{i=1}(n_a^il_b^i+n_b^il_a^i).
$$
The sign of $I$ signifies the chirality of the particle, with a
negative intersection number corresponding to a left-handed fermion.

D6-branes can also intersect with their images. Naively one might
assume that strings extended from a stack $a$ to stack $a'$ would
give particles transforming as $(N_a/2,\overline{N_a/2})$, but the
orientifold projection leads to two index symmetric and
antisymmetric tensor representations of $U(N_a/2)$.  The
intersection number between a stack and its image is given by
$$
I_{aa'}=[\Pi_a][\Pi_{a'}]=-2^{3-k}\prod^3_{i=1}(n_a^il_a^i).
$$
However, massless strings will also stretch between a stack of
branes and image at the orientifold planes, and so we must take into
account the intersection
$$
I_{aO6}=[\Pi_a][\Pi_{O6}]=2^{3-k}(-l_a^1l_a^2l_a^3+l_a^1n_a^2n_a^3+n_a^1l_a^2n_a^3+n_a^1n_a^2l_a^3).
$$
The final result for the net number of symmetric and anti-symmetric
representations is found by anomaly cancellation:
$$
{1 \over 2}(I_{aa'}-{1 \over 2}I_{aO6}) \hspace{.2in} {\rm symmetric}
$$
$$
{1 \over 2}(I_{aa'}+{1 \over 2}I_{aO6}) \hspace{.2in} {\rm antisymmetric.}
$$
The resulting chiral spectrum is listed in table \ref{dbranespectrum}.

\begin{table}[!h!t!b]
\begin{center}
\begin{tabular}{|c|c|c|}
\hline \hline
Sector &Representation &Multiplicity \\
\hline
&&\\
$ab+ba$ & $(N_a/2,\overline{N_b/2})$ & $
I_{ab}=2^{-k}\prod^3_{i=1}(n_a^il_b^i-n_b^il_a^i)$  \\
&& \\ 
$ab'+b'a$ &$(N_a/2,N_b/2)$&$
I_{ab'}=-2^{-k}\prod^3_{i=1}(n_a^il_b^i+n_b^il_a^i).
$ \\ 
&&\\ 
$aa'+aO6$& $(N_a \otimes N_a)_s$ &$ {1 \over 2}(I_{aa'}-{1 \over 2}I_{aO6}) $\\ 
& $(N_a \otimes N_a)_a$ &${1 \over 2}(I_{aa'}+{1 \over 2}I_{aO6})$  \\
&&\\
\hline \hline
\end{tabular}
\end{center}
\caption{Chiral Spectrum from Intersecting D6-branes.}
\label{dbranespectrum}
\end{table} 

A consistent supersymmetric theory must satisfy both tadpole and
supersymmetry constraints. Cancelation of RR tadpoles follows from
the cancellation of D6-brane and O6-plane charge, which implies
$$
\sum_a N_a [\Pi_a]+\sum_a N_a [\Pi_{a'}] -4[\Pi_{O6}]=0.
$$
To preserve supersymmetry, each D6-brane must be related to the
orientifold plane by an $SU(3)$ rotation.  Because the D6-branes are
taken to be products of one-cycles, each cycle will lie at some
angle $\theta_i$ with respect to the horizontal direction in the
$i$th torus.  The condition
$$
\theta_1+\theta_2+\theta_3=0 \ {\rm mod} \  2 \pi
$$
ensures that the total angle of rotation is an element of $SU(3)$.
The angles $\theta_i$ can be expressed in terms of the wrapping
numbers as
$$
\sin \theta_i={2^{-\beta_i}l^iR^i_2 \over L^i(n^i,l^i)}, \hspace{.2in} \cos \theta_i = {n^i R^i_1 \over L^i(n^i,l^i)},
$$
where $R^i_1,R^i_2$ are the radii of the horizontal and vertical
directions of the $i$th torus, and
$L^i(n^i,l^i)=\sqrt{(2^{-\beta_i}l^iR^i_2)^2+(n^iR^i_1)^2}$ is the
total length of the one-cycle on the $i$th torus.

\subsection{Search Strategy and Results}

In \cite{CPS}, Cveti\v{c}, Papadimitriou, and Shiu (CPS) have
performed an extensive search for $\mathcal{N}=1$ supersymmetric
three-family $SU(5)$ Grand Unified Models using the above
construction.  Therefore, we have used their work as the starting
point for our search for the Pentagon model.  In this section we
will discuss what adjustments must be made to the CPS models in
order to accommodate the inclusion of the Pentagon $SU(5)$ gauge
group and our required matter multiplets.  Based on simple
assumptions that these adjustments lead us to, the existence of the
Pentagon Model is ruled out.  In particular, the number of stacks
required to obtain the Pentagon spectrum introduces a problem with
the complex moduli, and the simplest solution to this problem is not
consistent with both tadpole and supersymmetry constraints. Relaxing
these assumptions leads to models that are far more complicated and
which must be evaluated on a case-by-case basis. Thus, while the
construction of consistent models in the context of the Pentagon is
not a forbidden possibility, it is left to future research.


We are looking to build a low energy phenomenological model that is
`at least as good' as the various CPS models, but with a few
additional requirements.  The CPS models are all four-dimensional
chiral models with N=1 SUSY constructed from IIA orientifolds on
$T^2 \times T^2 \times T^2 / (Z_2 \times Z_2)$.  They satisfy
consistency conditions (tadpole cancellation), preserve
supersymmetry, and contain three generations of $SU_{GUT}(5)$ matter
(or to be more precise, they all have 3 generations of the ${\bf 10}_a$ representation of the $SU_{GUT}(5)$, but with
varying number {\bf 5} fundamental representations).  These models also
have various phenomenological challenges, including the existence of
substantial numbers of chiral exotics as well as issues with the
Higgs fields and Yukawa couplings.  

We are willing to accept these
shortcomings for the present purpose, but there are other
requirements that must be satisfied to reproduce the low energy
spectrum of the Pentagon.  Primarily, we require the existence of an
additional stack of D-branes to give the $SU_{P}(5)$ of the Pentagon
model, and total topological intersection number zero between this
stack and the stack of the GUT $SU(5)$ (this is because of the
vector-like nature of the Pentaquarks, which transform as either
$(5,5)$ or $(5,\bar{5})$ plus c.c. under $SU_{P}(5) \times
SU_{GUT}(5)$).  This last requirement is satisfied by having the two
stacks parallel on the first $T^2$ (by choice), but we wish to
impose the additional constraint that the intersection number equal
one on the remaining two Torii, so that there is only a single point
at which the vector-like Pentaquarks may arise, thereby prohibiting
additional unwanted generations of the Pentaquarks which could be
disastrous for coupling unification and possibly introduce Landau
poles.  We also assume that the two parallel stacks on the first $T^2$ are actually lying right on top of each other, and further that they lie parallel to the orientifold plane.  The first of these ensures that the pentaquarks remain massless, and the latter that they have no intersections with their orientifold images\footnote{Of course, the $SU(5)$ stacks and their images will still be parallel, so their positions must be fixed at positions on the first torus such that string states stretching between a brane and its image are massive, i.e. non-zero distance.} (which would lead to exotic pentaquark-like fields charged under both $SU(5)$s).  See figure 4.1.  Finally, we would like to have two $U(1)$ stacks of D6-branes, the intersections of which would provide the singlets of the Pentagon.  The desired particle content is summarized in table \ref{pentcontent}.

\begin{figure}
 \begin{center}
\resizebox{\textwidth}{!}{\includegraphics{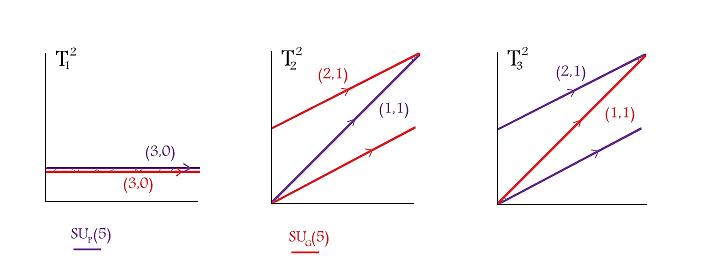}}
 \caption{Geometeric Requirement for the Pentaquarks.  Total intersection number is zero, with only a single intersection point (at the origin) in the second and third $T^2$.}
 \end{center}
\label{pqgeom}
\end{figure}

\begin{table}[!h!t!b]
\begin{center}
\noindent \begin{tabular}{|ccc|}
\hline \hline
Stack & Gauge Group & D6-brane Stack \\
\hline
a: & Pentagon $SU(5)$ & $N_a=10$ \\
b: &SM GUT $SU(5)$ & $N_b=10$ \\
c: &$U(1)$ & $N_c=2$ \\
d: &$U(1)$ & $N_d=2$ \\
\hline
\end{tabular} 
 \vspace{.1in}
\nolinebreak \hskip 2mm \begin{tabular}{|c|c|}
\hline 
 $SU_P(5) \times SU_{GUT}(5)$& Topological Intersection \\
Particle Representation & (Multiplicity) \\
\hline
 $(5,\bar{5})+(\bar{5},5)$ & $I_{ab} = 0$\\
 $(1,10_a)$ & ${1 \over 2} (I_{bb'} + {1 \over 2} I_{bO6}) =3 $ \\
 $(1,\bar{5})$& $I_{bc}+I_{bc'}+I_{bd}+I_{bd'}=-3$\\
Singlets & $I_{cd}+I_{cd'} \neq 0$\\
\hline \hline
\end{tabular}
\end{center}
\caption{Summary of Pentagon model D6-brane content and corresponding topological intersection numbers.} \label{pentcontent}
\end{table}

Let us begin by briefly listing the constraints relevant to
our criteria.  Following CPS, we define the parameters
$$
A_a=-n_a^1 n_a^2 n_a^3,\   B_a=n_a^1 l_a^2 l_a^3,\  C_a=l_a^1 n_a^2 l_a^3, \  D_a=l_a^1 l_a^2 n_a^3
$$
$$
\tilde{A}_a=-l_a^1 l_a^2 l_a^3, \   \tilde{B}_a=l_a^1 n_a^2 n_a^3, \  \tilde{C}_a=n_a^1 l_a^2 n_a^3, \,  \tilde{D}_a=n_a^1 n_a^2 l_a^3.
$$
Then the tadpole cancellation conditions can be rewritten as
$$
-16=-2^k N^{(1)}+ \sum_a N_a A_a = -2^k N^{(2)}+ \sum_a N_a B_a
$$
$$
=-2^k N^{(3)}+ \sum_a N_a C_a =-2^k N^{(4)}+ \sum_a N_a D_a.
$$
The $N^{(i)}$ correspond to the number of `filler branes' wrapping
the $i$th orientifold plane, which is for our purposes arbitrary and
can be used to reduce the total tadpole charge to the desired -16.
They always contribute negatively, so our biggest obstacle will be
to ensure that the sum of the charges from our stacks be large
enough.  The number of branes in each stack determines the gauge
group, so in our case we want $N_a=10, N_b=10, N_c=2, N_d=2$.  The
extra factor of two is due to the orientifold projection.  This
allows us to write out the tadpole constraints more explicitly:
$$
10 A_a + 10 A_b + 2 A_c + 2 A_d = -16 + 2^k N^{(1)}
$$
or to simplify things
$$
10 A_a + 10 A_b + 2 A_c + 2 A_d > -16
$$
and similarly for B, C, D.

The supersymmetry constraints must be satisfied for each stack
individually. The condition $\theta_1+\theta_2+\theta_3=0$ mod $2
\pi$ is equivalent to $\sin(\theta_1+\theta_2+\theta_3)=0$ and
$\cos(\theta_1+\theta_2+\theta_3)>0$, which can be rewritten in
terms of our new variables as
$$
x_A \tilde{A}_a +x_B \tilde{B}_a +x_C \tilde{C}_a +x_D \tilde{D}_a =0
$$
$$
A_a/x_A + B_a/x_B + C_a/x_C +D_a/x_D <0
$$
with similar expressions for stacks b,c,d.  The $x_A,x_B,x_C,x_D$
are related to the complex moduli of the tori $\chi_i=(R_2/R_1)_i$.
Only three are independent (for simplicity we can set $x_A=1$), and
each must be positive.  While we are free to adjust the moduli, each
stack of branes introduces a new constraint.  Thus generically three
stacks of branes completely fix the three moduli of the tori, and
so there is no freedom in adding a fourth stack of branes.  For this
reason, CPS only consider configurations with up to three
non-trivial stacks, and this is a significant problem which must be
addressed in our model.

CPS classify the possible brane wrapping configurations into four
types, based on the number of tori in which the stack of branes is
parallel to one of the orientifold planes (i.e. the number or $n$s
or $l$s equal to zero).  Type I has 3 zeros and so is completely
parallel to one of the orientifold planes, these are the so-called
`filler branes'.  Type II has two zeros, there are no SUSic
configurations with two zeros.  Type III has one zero, so a type III
stack is parallel to the orientifold plane in one of the three
tori.  In this case, exactly two of $A,B,C,D$ and two of
$\tilde{A},\tilde{B},\tilde{C},\tilde{D}$ are zero.  Without loss of
generality we can choose $n^1=0$, so that
$A=B=\tilde{C}=\tilde{D}=0$, $CD=-\tilde{A}\tilde{B}$, and imposing
the SUSY conditions we find
$$
C<0,\ D<0, \ \tilde{A}\tilde{B}<0,\ x_B=-\tilde{A}/\tilde{B}.
$$
Finally, type IV has no zeros, and so $A \tilde{A}=B \tilde{B}=C
\tilde{C}= D \tilde{D}= {\rm constant} \neq 0$.  Also, SUSY requires that
only one of $A,B,C,D$ is positive, and
$$
x_A/ {A}_a +x_B /{B}_a +x_C /{C}_a +x_D /{D}_a =0.
$$

Assuming a maximum of three non-trivial stacks of branes, CPS show
that the $SU_{GUT}(5)$ brane stack must be type III, and taking
$n_a^1=0$ they find $C_a=-1, D_a<-4$ with $k=1$ or $k=2$.  Tadpole
conditions then require a second stack with $D_b>0$ and must
therefore be type IV, with $A_b,B_b,C_b$ negative. We still have the
freedom to add a third stack of either type III or type IV, with the
requirement $C_c \leq 0$.

In our case, since we are interested in four stacks, we would like
two of the stacks to obey the same equation for the moduli, i.e. to have equal angles with respect to the orientifold plane.  The
natural choice then seems to be the GUT and Pentagon stacks since
they must be parallel anyway.   Our strategy then will be to
consider the two $SU(5)$ stacks to be parallel in the first torus,
but with the orientation of one stack in the second torus parallel
to that of the other stack in the third and vice versa (see figure 4.1).  So for example, two stacks with the winding numbers
$$
SU_{P}(5):    (0,3) \times (1,1) \times (2,1)
$$
$$
SU_{GUT}(5):  (0,3) \times (2,1) \times (1,1)
$$
would obey the same moduli equations; that is, the supersymmetry
constraints on these stacks fix only one of the three moduli because
the total angle of the stacks are the same. You will notice that
this configuration has the two stacks parallel in the first torus
(and therefore total topological intersection number of zero), while
the number of intersection points in the last two tori is one as we
would like, and if you calculate the number of chiral fields in the
antisymmetric representation of the $SU_{GUT}(5)$ you will find the
desired three families.  In this example we would have
$C=-3,D=-6,\tilde{A}=3,\tilde{B}=-6, x_B=1/2$ for the Pentagon stack
(stack $a$), and $C=-6,D=-3,\tilde{A}=3,\tilde{B}=-6, x_B=1/2$ for the
GUT stack (stack $b$).   Unfortunately, this model is just one example of an entire class of similar models which suffer an
incompatibility between the tadpole constraints and the SUSY conditions, as follows.

We have mentioned that stacks $a$ and $b$ must be parallel to the orientifold plane in order to avoid pentaquark-like exotics.  This follows from the fact that we have demanded the number of intersection points in the second and third tori to be exactly one.  If the two stacks were {\it not} parallel to the O6-plane in $T^2_1$, stack $a$ would certainly intersect with the image of stack $b$ in that torus (with the reverse being true as well).  We might have hoped that the topological intersection could still be zero if there is a cancellation $(n_a^il_b^i+n_b^il_a^i) = 0, i=2 \ {\rm or} \ 3$, but this cannot be true if $(n_a^il_b^i-n_b^il_a^i) = 1, i=2, 3$.  Since $n,l$ are integers, there is no way to add two numbers to get one and subtract them to get zero.  Thus, both the Pentagon and GUT stacks must be type III.

However, if this is true, we cannot
simultaneously satisfy the tadpole and supersymmetry conditions.
Let us enumerate some of the requirements on a and b if they are
both to be type III.  First, since the number of antisymmetric
tensor representations for stack a is given by $I_{aa'}+1/2 I_{aO6}=3$, and for
type III $I_{aa'}=0, I_{aO6}=2^{3-k}(\tilde{A}+\tilde{B})$, we find
that either $k=1$ with $\tilde{A}+\tilde{B}=3$ or $k=2$ with
$\tilde{A}+\tilde{B}=6$, and in each case $\tilde{A}\tilde{B}<0$ as
before.  Second, if we are to have the two stacks parallel in the
first torus (by choice), we must have either $n_a^1=n_b^1=0$ or
$l_a^1=l_b^1=0$ in order for them to satisfy the same moduli
equations.  We will choose the former for convenience.   Finally,
the requirement that we have only one intersection between stack a
and b in the last two tori implies $n_a^2 l_b^2 - l_a^2 n_b^2 =
1,n_a^3 l_b^3 - l_a^3 n_b^3 = 1$. These conditions are solvable yet
very confining, the simplest solution of which was given in the
example above.

The problem then is this.  We know that the values of $C$ and $D$
are both negative integers for both stacks a and b, and in fact
$C_a=D_b,C_b=D_a$ for the type of configurations where stack a and b
obey the same moduli equations as suggested above. This alone
already implies that $C_a+C_b=D_a+D_b<-2$, but if the stacks are to
satisfy the requirements of the previous paragraph the statement is
more severe with $C_a+C_b=D_a+D_b<-9$.  Recall that the tadpole
condition instructs us to multiply these factors by the number of
membranes in each stack, which again is $N_a=N_b=10$, so that at
best we have $-90 +2C_c+2C_d>-16$ and similarly for $D$.  In other
words, either $C_c$ or $C_d$ as well as $D_c$ or $D_d$ must be large
and positive.  This immediately rules out the possibility that
stacks c and d are type III, because we know that for type III $C$
and $D$ are less than or equal to zero.  For type IV stacks, only
one of $A,B,C,D$ can be positive and still satisfy SUSY conditions,
so our only hope is that say $C_c >0$ and $D_d>0$ and that both
values are large.  Unfortunately, even this doesn't work.  If
$C_c>0$ then $D_c$ will contribute negatively to the $D$ tadpole
conditions, so of course we must require $|D_d|>|D_c|$, similarly
$|C_c|>|C_d|$.  This then leads to a problem with the moduli.  For
stacks c and d we have
$$
A_c+x_B/B_c+x_C /C_c+x_D/D_c=0
$$
$$
A_d+x_B/B_d+x_C /C_d+x_D/D_d=0.
$$
We have already solved for $x_B$ previously (it is positive), so the
the first two terms in each of these equations sum to a negative
number.  Multiply the equations by $C_c D_c$ and $C_d D_d$
respectively, and we can rewrite these as
$$
|D_c| x_C - |C_c| x_D = P
$$
$$
-|D_d| x_C + |C_d| x_D = Q,
$$
where $P,Q$ are positive numbers.  Summing the two equations we find
$$
(|D_c|-|D_d|)x_C + (|C_d|-|C_c|)x_D = P+Q
$$
implying that at least one of $x_C$ or $x_D$ is negative. This
argument is analogous to the arguments in CPS given to forbid case
(iv), $k=2$ and case (i), $k=3$ for type IV branes and case (i) for
type III branes.  Therefore, stacks a and b cannot be required to solve the same moduli equation.  In fact, the argument is even stronger:  stacks a and b cannot both be type III.  This implies the existence of pentaquark-like exotics.   

What if we relax this last requirement, i.e. not demanding stacks a and b to be type III?  In order for two stacks to
solve the same moduli equations, they must both be of the same type,
so the question leads us to consider the compatibility of two stacks
of type IV branes.  The answer in this case is simple, and in fact
applies regardless of the gauge groups supported on the stacks or
their intersection number.  As we have seen, the moduli equations
for type IV can be written
$$
A+x_B/B+x_C /C+x_D/D=0,
$$
and if any two stacks are to both obey the same equation we must
have $A_1=A_2,B_1=B_2,C_1=C_2,D_1=D_2$.  In this case, as far as the
tadpole conditions and supersymmetry constraints are concerned, we
can then just consider these two stacks as a single stack with
$N=N_1+N_2$ and $A=2A_1$, etc.  But we already know the requirements
for a consistent model with three stacks.  In particular, the GUT
stack must be type III by the requirement of correct family
multiplicity.  Therefore, the possibility of the $SU_{GUT}(5)$ and
$SU_{P}(5)$ stacks solving the same moduli equation is completely
excluded.  If we wish any other combination of type IV branes to
solve the same moduli equation, the problem again reduces to the
discussion of three stacks\footnote{This also provides an alternative argument proving the existence of pentaquark-like exotics in these models.  If stacks a and b are both type III, they cannot be required to solve the same moduli equations.  Because two stacks cannot solve the same moduli equation if they are of different types, this responsibility falls on the type IV stacks c and d.  As argued, we can then consider these as a single stack.  But we know that the Pentagon does not exist in a model with three stacks, two of which are type III.  At least one of stacks a,b, then, must be type IV.}. 

We are left with two possible approaches.  The first would be to
return to the possibility of constructing a consistent model with
three stacks of D6-branes.  We would then have to either assume that
two of the stacks exactly obey the same moduli, supersymmetry, and
tadpole equations (as suggested above), or to abandon one of the
$U(1)$ stacks and argue that the Pentagon singlets arise from
another mechanism.  However, the question of $SU(5)$ GUT theories
with three stacks of branes was exactly the subject of the CPS
search.  The GUT stack must be type III, and a second stack must be type IV.  In their paper, they have listed all 149 possible solutions
for models with a third stack of type III, none of which contain a
second $SU(5)$ gauge group.  Thus, the $SU_P(5)$ and any $U(1)$
factors must arise from stacks of type IV.  According to CPS, this
would require a very extensive search that would have to be
conducted on a case by case basis, a search that CPS didn't endeavor
to attempt.  In any case, we believe it likely that a proof could be
constructed to show that this possibility is inconsistent due to the
severity of the constraints imposed by adding a second $SU(5)$ gauge
group.  This will be the subject of a future investigation.

The second possible approach is to allow a different combination of
stacks to obey the same moduli equation.  As we have argued, both
these stacks would have to be type III or else the problem again
reduces to a question of three stacks.  We know that at minimum one
of the stacks has to be type IV to satisfy the tadpole conditions,
and this stack would likely have to sustain the $SU_P(5)$.  If we
make this assumption, we would have to find a non-trivial
combination of two stacks with $N_a=10$ and $N_b=2$ which satisfy
the same moduli equation.  The argument would then parallel that of
two type III stacks given above, but with some of the assumptions
made there relaxed.  If such a search were to fail, we would have to
completely abandon the hope that two of our four stacks exactly
solve the same moduli equation, and would be forced to find a system
of equations in which the fourth stack obeys an equation which is a
non-trivial linear combination of the other three.  We have not yet
found a strategy for systematically attacking this problem.

In any case, we know that these models will contain many undesired chiral exotics.  CPS have shown that the existence of $15_{sym}$ representations are unavoidable in models with $10_a$s.  We have further demonstrated that our models will contain chiral pentaquark-like exotics, charged as $(5,5)$ under the Pentagon and GUT gauge groups.  These particles are surely phenomenologically untenable.

Furthermore, though our search for the particle content of the Pentagon has proven somewhat inconclusive to this point, we find it
likely that no self-consistent solution exists.  The major culprit
for this difficulty seems to be the tadpole constraints imposed by
requiring two $SU(5)$ gauge groups.  Generally speaking, the larger
the gauge group, the more negatively a stack will contribute to the
RR-charge.  This fact, combined with the requirement that we find
one (and only one) vector-like pair of pentaquarks, seems to be too
great an obstacle to overcome.  This leads us to believe that these
constraints might be softened if we were to consider searching for
the particle spectrum of the Pyramid model, for which we would need
an $SU(3)^4$ gauge group arising from four stacks of $N=6$
D6-branes.  At the current time we are in the preliminary stages of
such a search, and the approach seems promising.

\section{M-theory on $G_2$ Manifolds}

Because they carry no fluxes or additional charge sources, D6-branes
and O6-planes are seen to be pure geometrical artifacts in the
strong coupling limit.  This suggests that one might consider an
M-theory description of chiral particles arising at points in the
manifold.  In particular, we are led to believe that $\mathcal{N}=1$
globally supersymmetric type IIA intersecting D6-brane models lift
up to eleven dimensional M-theory compactifications on singular
$G_2$ manifolds \cite{Cvetic:2001kk}.  D6-branes and O6-planes wrap
smooth supersymmetric three cycles in the IIA compactifications, and
one fibers each of these by a suitable noncompact hyperk\"{a}hler
four-manifold to obtain the $G_2$ holonomy space. In the M-theory
language, these are codimension four ADE-orbifold singularities
spanning three cycles in the $G_2$ compactification manifold, and
must be ALE (asymptotically locally Euclidean) spaces. $N$
overlapping D6-branes correspond to an $A$-type ALE singularity,
$D$-type singularities arise for D6-branes overlapping O6-planes.
Chiral fermions exist at isolated co-dimension seven singularities,
which would correspond to the $G_2$ lift of the intersection points
of D6-branes and O6-planes in the IIA picture.  Just as in the IIA
constructions, family replication is given by the number of these
singular points in the manifold.  When a point on the manifold
shrinks to a conical singularity, the symmetry supported along that
fiber will be enhanced at the singularity.  To determine the chiral
representations arising there, we decompose the adjoint of the group
associated with higher symmetry with respect to that of the
lower\cite{Mtheory}.

Specifically \cite{Acharya}, we will obtain chiral fields in the
representation $R$ of group $G$ if at certain points on the manifold
the $G$ singularity is enhanced to a group $\hat{G} = G \otimes
U(1)$.  Away from these points, the Lie algebra of $\hat{G}$ will
decompose as
$$
\hat{g} \rightarrow g \oplus o \oplus r \oplus \bar{r}
$$
where $g$ and $o$ are the Lie algebras of $G$ and $U(1)$, $r$
transforms as $R$ (and of charge 1 under $U(1)$), and $\bar{r}$ the
complex conjugate.  However, Acharya and Witten have shown that the
net number of chiral zero modes is one, meaning that only either $r$
or $\bar{r}$ will appear in the low energy theory (depending on how
the chirality is fixed)\footnote{The discussion is complicated in
the case of semi-simple $G$, see \cite{Bourjaily:2008ji}}.  The
group $G$ need not be simple, it may be any semi-simple product of
the groups obtained by deleting one node from the Dynkin diagram of
$\hat{G}$.  The representations found at a particular singularity
will not always be free from anomalies; however one can show that
when this is the case there must exist another point (or set of
points) elsewhere on the manifold supporting particles which render
the theory anomaly free.

\begin{table}[!h!t!b]
\begin{center}
\begin{tabular}{|l|l|}
\hline
Resolution & Chiral Representation \\
\hline
$SU(N+1) \rightarrow SU(N)$ & $[N]$ \\
$SU(N+M) \rightarrow SU(N) \times SU(M)$ & $[N,\bar{M}]$ \\
$SO(2N) \rightarrow SU(N)$ & $[N \otimes N]_a$ \\
$SO(2(N+1)) \rightarrow SO(2N)$ & $[2N]$ \\
\hline
\end{tabular}
\end{center}
\caption{Resolutions at A and D type singularities on manifolds of $G_2$ holonomy.}
\label{G2breaking}
\end{table}

For $A$ and $D$ type singularities we are led to the representations
listed in Table \ref{G2breaking}.  Note that these are in agreement
with the picture we have from IIA intersecting D6-brane models.  For
two three-cycles intersecting at a singular point in the $G_2$
manifold, supporting gauge groups $SU(N)$ and $SU(M)$ respectively,
we are left with chiral fermions in the bifundamental representation
$(N,\bar{M})$. This is just as we would expect from the intersection
of two stacks with $N$ and $M$ D6-branes.  Similarly, the resolution
$SO(2N) \rightarrow SU(N)$ leads to the antisymmetric representation
of $SU(N)$, corresponding to particles which would be found at the
intersection of a stack of $N$ D6-branes with an O6-plane.  However,
the parallel should not be taken too literally.  Unlike the IIA
picture, three cycles in a seven-manifold do not generically
intersect, so the existence of multiply charged particles will only
be found in specially constructed geometries.

Can the Pentagon model be embedded in a $G_2$ compactification of
M-theory?  To answer this, we must find a 4-d theory with an
$SU_P(5) \times SU_{SM}(5)$ gauge group and chiral fermions in the
representations $(5,\bar{5})$ and $(\bar{5},5)$ (or possibly $(5,5)$
and $(\bar{5},\bar{5})$), $3 \times (1,10)$, $3 \times (1,\bar{5})$,
a pair of Higgs $(1,5),(1,\bar{5})$, and a singlet field $S$ (which
we will ignore for the moment--we might assume it arises as some
modulus of the geometry); all of which arise at singularities in the
$G_2$ manifold. There is no single point that can sustain a symmetry
which unfolds to the $SU(5) \times SU(5)$ gauge group plus all of
the desired matter, so our model will necessarily have to be a
patchwork of fields lying at different points in the manifold.  This
is not necessarily a problem; such a geometry would surely be less
generic, but it could help explain the large family hierarchies of
the standard model.  The proximity of matter multiplets to Higgs
fields would vary from point to point, creating a natural hierarchy
in the Yukawa couplings.

It is clear that the desired components of our model can be derived
in this construction, and our search for such a model in the IIA
context points to the answer. Consider a three-cycle in the $G_2$
manifold sustaining an $SU(5)$ ADE-orbifold singularity.  If at
certain points along this cycle we find a conical singularity, the
symmetry will be enhanced.  If the enhanced gauge group is $SU(6)$,
the symmetry will unfold as $SU(6) \rightarrow SU(5) \oplus o \oplus
5 \oplus \bar{5}$.  Let us suppose the zero modes are the
$\bar{5}$s, and that there are four of such points.  Anomaly
cancellation ensures that elsewhere on the manifold we can find
representations of opposite chirality, but let us assume we find
only one such point (leaving us with one 5 representation).
Let the additional anomalies be canceled by $10_a$ representations,
arising at singularities where the symmetry has been enhanced to
$SO(10)$, i.e. $SO(10) \rightarrow SU(5) \oplus o \oplus 10 \oplus
\bar{10}$.  Now assume that there exists an additional three-cycle
on the manifold supporting a new $SU(5)$ symmetry, and that these
two cycles somewhere intersect at a point.  If this special point
happens to lie at a conical singularity, the symmetry will be
enhanced to $SU(10)$, which will resolve as $SU(10) \rightarrow
SU(5) \times SU(5) \oplus o \oplus (5,\bar{5}) \oplus (\bar{5},5)$.
Here we will find the pentaquarks, and elsewhere on the manifold we
must find the anti-pentaquarks to cancel anomalies.  Thus, this
$G_2$ manifold will support two $SU(5)$ symmetries as well as the
entire matter content of the Pentagon model, with no exotics (see
Table \ref{G2model}).

\begin{table}[!h!t!b]
\begin{center}
\begin{tabular}{|c|c|c|c|}
\hline \hline
Location & Supported Gauge Group & Enhanced Singularity & Matter Content \\
\hline
Three-Cycle 1& $SU(5)$ & $5 \times SU(6)$ & $4 \times \bar{5}$ \\
&&& $1 \times 5$ \\
&  & $3 \times SO(10)$ & $3 \times 10_a$ \\
\hline
Three-Cycle 2 & $SU(5)$&&\\
\hline
Intersection& $SU(5) \times SU(5)$ & $2 \times SU(10)$ & $[5,\bar{5}]$ \\
Points&&& $[\bar{5},5]$ \\
\hline \hline
\end{tabular}
\end{center}
\caption{M-theory model containing the Pentagon spectrum.}
\label{G2model}
\end{table}

Clearly, such a model will be highly non-generic.  We may desire a
model which sustains (at minimum) a pervasive symmetry $G=SU(5)
\times SU(5)$ throughout the entire manifold, allowing the gauge
group to be defined throughout the bulk.  However, such a
requirement complicates the model significantly.  As we have seen,
the pentaquarks can be found at points where the symmetry is
enhanced to $SU(10)$, but this leaves no freedom to derive the
chiral fields of the model\footnote{There may be an exception to this statement.  We are currently investigating the possibility of singularities enhanced to $SU(6) \times SU(5)$ or $SO(10) \times SU(5)$.  However, it is unclear to us at the present moment whether these fields will be charged under the second $SU(5)$ or have other undesirable $U(1)$ charges.}.  On the other hand, the $SU(5) \times SU(5)$ gauge group can be obtained from Higgsing an $SU(10)$ with
Wilson lines, so let us have the $G_2$ manifold support a pervasive
$G = SU(10)$.

One possibility is that $G$ uplifts to $\hat{G}=SU(11)$, in which
case $\hat{g} \rightarrow su(10) \oplus o \oplus 10 \oplus
\bar{10}$.  With Wilson lines, the $10$ will further decompose as
$(1,5) + (5,1)$.  Five of these points (with the proper chirality)
will provide us with the standard model $3 \times (1,\bar{5})$ as
well as the Higgs fields.  However, we are left with some
potentially undesirable particles, namely the four $(\bar{5},1)$s
and the $(5,1)$.  We can imagine that the $(5,1)$ will mass up with
one of the $(\bar{5})$s, but we are still left with $3 \times
(\bar{5},1)$.  These particles are not necessarily problematic, as
they have no standard model quantum numbers or interactions.  In
fact, they may have the potential of providing us with a dark matter
candidate.  For this to be possible, we would need to find
$\lambda^{ijk}$s (where $\lambda^{ijk} 10^i \bar{5}^j \bar{5}^k$)
such that the $U(3) \times U(3)$ flavor symmetry is broken to a
conserved $U(1)$ with $Tr [T^3] = 0$.  For now, let us just assume
that these fields pose no phenomenological problems for the model.

If at another point we find $\hat{G} = SO(20)$, we would be left
with $SU(10) + o + 45 + \bar{45}$ where the $45_a$ is the
antisymmetric tensor representation of $SU(10)$.  Higgsing the
$SU(10)$, the 45 decomposes as $(5,5) + (1,10) + (10,1)$ under
$SU_P(5) \times SU_G(5)$, leaving us with candidates for the
pentaquarks and the standard model $10_a$.  However, in order to
have three standard model generations there must be three separate
points on the manifold with a $\hat{G} = SO(20)$ singularity.  Thus
we obtain (one half of) the desired pentaquarks (charged as $(5,5)$
as opposed to $(5,\bar{5})$), as well as the $3 \times (1,10)$ of
the standard model; we also are left with an undesirable two
additional copies of $(5,5)$ and with three $(10,1)$.  The latter of
these is possibly interesting (as discussed above), but the former
spells disaster.  Fortunately, with proper $Z_4$ R-charge
assignments (two of the three (5,5) with R-charge 0 and one with
charge 2), we might imagine one pair massing up and leaving us with
just the single generation of pentaquarks.  Furthermore, this $Z_4$
could conceivably ensure that the $10$s do not gain mass.

Of course, so far this model is not anomaly free. This is perhaps
fortunate, because we are guaranteed to find the vector-like pair
for the pentaquarks elsewhere on the manifold.  We might be tempted
to think the anomalies of the numerous 5 and 10 representations
exactly cancel (there are equal numbers of $\bar{5}$s with $5 +
10_a$s), but there would then be no way to construct the
anti-pentaquarks $(\bar{5},\bar{5})$.  Thus we are forced to
consider an additional three points elsewhere in the manifold
supporting $\hat{G} = SO(20)$ singularities giving rise to the
conjugate pairs.  But this would lead to vector-like pairs $10 +
\bar{10}$, which is clearly unacceptable.  This would also force us
to include additional 5 representations to cancel the anomalies of
the $\bar{5}$s, and whether or not these pairs gain mass or remain
light, this is certainly phenomenologically untenable.

One potential solution to this problem would be to soften our
requirement for a pervasive $SU(10)$ gauge symmetry to a pervasive
$SU(5) \times SU(5)$, but containing an entire fiber with enhanced
symmetry $SU(10)$.  The pentaquarks of this model would arise at
points away from this fiber, but also enhanced to an $SU(10)$
symmetry, and unfolding as $SU(10) \rightarrow SU(5) \times SU(5)
\oplus o \oplus (5,\bar{5}) \oplus (\bar{5},5)$ plus its complex
conjugate.  Along the fiber, certain points would have `worsened'
singularities with the enhanced symmetries $\hat{G} = SO(20), \
SU(11)$.  This would give rise to all of the desired components
listed above, with the anomalies of the $\bar{5}$s and $10_a$s
exactly canceling.  The $(5,5)$s arising from the $SO(20)$ conical
singularities would be exotics in this model.  However, if for every
$(5,5)$ representation there were an additional five singularities
with $SU(11)$ symmetry, each producing a $(\bar{5},1)+(1,\bar{5})$
representation, the anomalies of these undesired particles would
cancel.  Furthermore, with correct R-charges, we might hope that all
of these exotics gain mass.  Of course, this model seems no more
aesthetically viable than those without a pervasive symmetry listed
above.

We therefore believe that the best candidate for our model is a
$G_2$ manifold supporting two three-cycles with $SU(5)$ gauge
symmetry intersecting at exactly two points.  One of these cycles
will support the standard model GUT $SU(5)$, and there will be
points along this cycle at which the symmetry is enhanced to either
$SU(6)$ or $SO(10)$ giving rise to the standard GUT matter. The
intersection points must lie at special points on the manifold where
the K3 structure has an enhanced symmetry, such that we find $SU(10)
\rightarrow SU(5) \times SU(5) \oplus o \oplus (5,\bar{5}) \oplus
(\bar{5},5)$ at one intersection and the vector-like partners at the
other.  Though this type of geometry is certainly highly
non-generic, it arises naturally from a lift of type IIA
intersecting D6-branes.  In such a model, we expect to find chiral
particles in the bifundamental representation exactly at such points
where two three-cycles intersect.  Indeed, the fact that such a
construction is possible in the M-theory context suggests that we
might hope to find a consistent model of intersecting D6-branes.
Conversely, the difficulty we have found in constructing such models
might suggest that the existence of such an M-theory geometry is
dubious.  Unfortunately, the tools necessary to perform an explicit
calculation are unknown, and the existence of such a model remains
in question.

\section{Heterotic Orbifold Constructions}

Some of the most realistic phenomenological string models have been
produced in the framework of heterotic
orbifolds\cite{heteroticmodels}.  This presents us with a promising
approach toward discovering a phenomenologically viable low energy
model.  In this section we will briefly review the heterotic
orbifold construction\footnote{In addition to the references given
in \cite{heteroticmodels}, see \cite{APthesi} for further details},
and then proceed to discuss our search strategy and preliminary
results.

\subsection{Brief Review}

Heterotic string theory is a theory of closed strings, combining the
supersymmetric right moving string with the left moving bosonic
string.  As the (uncompactified) theory must exist in 10 space-time
dimensions, the extra 16 dimensions of the bosonic string are
interpreted as internal degrees of freedom.  To satisfy modular
invariance (and anomaly cancellation) these 16 left movers must live
on a 16 dimensional Euclidean even self-dual lattice, which we
choose to be the root lattice of $E_8 \times E_8$ (although the
$SO(32)$ root lattice has been shown to be of interest as
well \cite{NillesSO32}).

The low energy effective field theory will consist of those states
which survive to low energies.  In particular, any state with mass
at the string scale will not be observed at low energy, so we are
only concerned with string states of zero mass. Further, since the
physical heterotic string states are the direct product of the right
movers with the left movers, both the right-moving and left-moving
string states must be massless. Working in the light cone gauge, we
find that there is a total of 16 massless right movers, 8 in the NS
sector which transform as an $SO(8)$ vector and 8 in the R sector
transforming as the $SO(8)$ spinor.  When tensored with the left
movers, the vector representation will produce the boson of the 10
dimensional supersymmetric chiral fields, while the spinor gives its
fermionic superpartner.  The 8 bosonic left moving oscillators
corresponding to the space-time degrees of freedom create massless
states when acting on the left moving ground state; when tensored
with the right movers, these form the $\mathcal{N}=1$ supergravity
multiplet. Similarly, the 16 internal degrees of freedom bosonic
oscillators acting on the left moving ground state form the 16
uncharged gauge bosons of $E_8 \times E_8$ (and their superpartners)
when tensored with the right movers.  Finally, there are the
massless 240 + 240 charged gauge bosons (plus superpartners) of $E_8
\times E_8$, which come from the tensor product of the right movers
with those left moving states having internal momenta satisfying
$(p^I)^2 = 2$. This is exactly the condition for the root vectors of
$E_8$, ensuring that these states lie on the $E_8 \times E_8$ root
lattice. Altogether, the massless heterotic string states form a
ten-dimensional $\mathcal{N}=1$ supergravity theory with $E_8 \times
E_8$ gauge group.

As the Pentagon model is a four dimensional low energy effective
field theory, six of the space-time dimensions must be compactified.
The simplest way to achieve this is to wrap each of these extra
coordinates on a circle, which is topologically equivalent to
compactifying on the 6-torus $T^6$.  However, torus compactification
schemes in general do not lead to realistic models in four
dimensions.  In particular, the $SO(8)$ spinor of the 10 dimensional
heterotic theory compactified on $T^6$ gives a total of 4 gravitinos
in 4 dimensions, thus leading to $\mathcal{N}=4$ supersymmetry.  To
obtain a chiral theory with $\mathcal{N}=1$ supersymmetry, one may
compactify on an orbifold
$$
\mathcal{O}=T^6/\mathcal{P} \otimes T_{E_8 \times E_8}/\mathcal{G},
$$
where the space-time and internal degrees of freedom are
differentiated to admit a clear space-time interpretation. Formally,
an orbifold is defined to be the quotient of a torus over a discrete
set of isometries of the torus, called the point group
$\mathcal{P}$.  The simplest of these is the symmetric abelian
orbifold, where the point group is chosen to be the cyclic group
$Z_N$ with $N=3,4,6,7,8,12$.  The lattice on which $\mathcal{P}$
acts as an isometry will be the root lattices of semi-simple Lie
algebras of rank 6.  The space group $\mathcal{S}$ is defined to be
the point group $\mathcal{P}$ plus the translations given by these
lattice vectors, such that $T^6/\mathcal{P}=R^6/\mathcal{S}$.  The
action of the space group on the (complex) space-time degrees of
freedom can be written as
$$
Z^a \rightarrow e^{(2 \pi i v^a)} Z^a + n_{\alpha}e^{\alpha}, \ a=1,2,3
$$
where $v$, the generator of the discrete group $Z_N$, is called the
twist vector, and the $e^{\alpha}$ are the lattice vectors of the
root lattice spanning $T^6$.  Thus, two points on $R^6$ are
identified if they differ by the action of the space group.  Points
that are invariant under the action of the space group are known as
fixed points of the orbifold. To ensure that exactly one 4
dimensional space-time supersymmetry survives, $\pm v^1 \pm v^2 \pm
v^3 = 0$ mod 2 with none of the $v^a$ vanishing.

$\mathcal{G}$ is called the gauge twisting group.  Modular
invariance requires the action of the space group to be embedded
into the gauge degrees of freedom. This means that in general the
internal gauge group of the orbifold will be a subgroup of the $E_8
\times E_8$ gauge group of the uncompactified heterotic theory.  To
realize this embedding, the orbifold twist vector is associated with
a shift vector $V$ in the $E_8 \times E_8$ root lattice, while the
torus shifts $e^{\alpha}$ are embedded as shifts $W_{\alpha}$.
Since the $W_{\alpha}$ correspond to gauge transformations
associated with non-contractible loops, they are interpreted as
Wilson lines.  The action of the gauge twisting group $\mathcal{G}$
on the gauge degrees of freedom is
$$
X^I \rightarrow X^I +2 \pi (k V^I + n_{\alpha} W_{\alpha}^I).
$$
The combined action of $\mathcal{S} \otimes \mathcal{G}$ is known as
the orbifold group.

Not all gauge twists and discrete Wilson lines are physically
allowed.  Modular invariance automatically guarantees the anomaly
freedom of orbifold models. For the partition function to be modular
invariant, it must satisfy the following conditions:
$$
(V^2-v^2)=0 \ {\rm mod} \ 2
$$
$$
V \cdot W_{\alpha} = 0 \ {\rm mod} \ 1
$$
$$
W_{\alpha} \cdot W_{\beta}=0 \ {\rm mod} \ 1, \ \alpha \neq \beta
$$
$$
W_{\alpha}^2=0 \ {\rm mod} \ 2.
$$
These conditions are known as `strong modular invariance'.  In
reality one need only satisfy `weak modular invariance', where these
conditions are slightly relaxed:
$$
N(V^2-v^2)=0 \ {\rm mod} \ 2
$$
$$
NV \cdot W_{\alpha} = 0 \ {\rm mod} \ 1
$$
$$
NW_{\alpha} \cdot W_{\beta}=0 \ {\rm mod} \ 1, \ \alpha \neq \beta
$$
$$
NW_{\alpha}^2=0 \ {\rm mod} \ 2.
$$
However, if weak modular invariance is satisfied, we can in general
bring $V$ and $W_{\alpha}$ to a form which obeys strong modular
invariance by adding $E_8 \times E_8$ lattice vectors.  This has the
advantage of simplifying the projection conditions on physical
states.\footnote{There are exceptions to this rule, such as when
$V=0$.}  $N$ is the order of the orbifold (and of the cyclic group
$Z_N$).  Cyclic group multiplication rules require that $N$
successive rotations of the orbifold act as the identity $Nv = 0$
mod 1, and that $NV$ belongs to the $E_8 \times E_8$ lattice.

The gauge transformations are required to be a symmetry of the
system.  To calculate which states survive orbifolding, we must
consider the action these transformations have on the states with
right- and left-moving momentum.  Neither the shifts nor the twists
act on the oscillators.  The generator of translation is
$$
e^{ip \cdot X} |0 \rangle = |P \rangle,
$$
so a shift in the coordinate degrees of freedom acts as a phase
rotation on the states.  For the right movers,
$$
|q \rangle \rightarrow e^{2 \pi i q \cdot(kv)}|q \rangle
$$
and for the left movers,
$$
|P\rangle \rightarrow e^{2 \pi i p \cdot(kV+n_{\alpha}W_{\alpha})}|P\rangle.
$$
States that are invariant with respect to the orbifold group
transform trivially (with a phase of 1) under every element of the
group, i.e. for all $k,n_{\alpha}=0,...,N-1$.  Only invariant states are
consistent with the geometry of the underlying orbifold space; all
other states must be projected out.

The massless spectrum consists of all massless closed string states
consistent with the geometry of the orbifold.  This includes the
massless strings of the original heterotic theory which survive the
projection conditions, as well as additional new states which arise
due to the non-trivial geometry of the orbifold.  The former form
the untwisted sector and are free to move throughout the orbifold,
while the latter are known as twisted sector states and are confined
to the fixed points.

First consider the untwisted sector. As mentioned earlier,
orbifolding projects out three of the supersymmetries, and we are
left with an $\mathcal{N}=1$ supergravity multiplet (as well as
certain modulus fields which are not relevant to the current
discussion).  The 16 uncharged gauge bosons correspond to the Cartan
generators of the $E_8 \times E_8$ algebra.  By construction, the
gauge twists and Wilson lines must commute with the Cartan
subalgebra, thus all uncharged gauge bosons (and gaugino partners)
survive the orbifold projection. Furthermore, the rank of the
algebra can never be reduced by the shift embedding.  The charged
gauge bosons of the heterotic string give rise to both the unbroken
gauge group as well as charged matter states.  As these are states
with both right- and left-moving momenta, they transform under the
orbifold group as
$$
|q\rangle \otimes |P\rangle \rightarrow e^{2 \pi i (q \cdot (kv)+p \cdot (kV+n_{\alpha}W_{\alpha}))} |q\rangle \otimes |P\rangle .
$$
The momenta of the right movers are given by their $SO(8)$ weights:
$$
q=(\underline{\pm 1,0,0,0}) \hspace{.25in} {\rm bosons}
$$
$$
q=(\pm 1/2, \pm 1/2, \pm 1/2, \pm 1/2) \hspace{.25in} {\rm fermions}.
$$
The underline denotes that all permutations are included.  Only
states with an even number of minus signs are included for the
fermions.  Gauge bosons in four dimensions have two transverse
polarizations, and so require oscillators in the uncompactified
directions, i.e. $q=(\pm 1, 0, 0, 0)$ in common
notation.\footnote{In complex coordinates, the first component
refers to the uncompactified directions and the last three
components to the coordinates on the six-torus $T^6$.  The lightcone
coordinates are gauge fixed and are omitted.  Typically, the first
component is omitted when writing the twist vector, $v$, as it must
be zero.}  Similarly, the gaugino states must have the right movers
$q=\pm (1/2,1/2,1/2,1/2)$.  Thus, the right movers of the four
dimensional gauge bosons (and gauginos) are invariant under the
orbifold action, $q \cdot v =0$.  The left movers, then, must
satisfy
$$
p \cdot (kV) = 0 \ {\rm mod} \ 1
$$
$$
p \cdot (n_{\alpha}W_{\alpha}) = 0 \ {\rm mod } \ 1
$$
for all $k,n_{\alpha}$.  Not all of the charged gauge bosons of the
heterotic string will satisfy these conditions, so the gauge group
is broken; those that do survive (along with the 16 Cartan
generators) form the generators of the unbroken gauge group on the
orbifold.  However, there are additional states which satisfy
$$
q \cdot (kv)+p \cdot (kV+n_{\alpha}W_{\alpha}) = 0 \ {\rm mod } \ 1
$$
without fulfilling $q \cdot v =0$.  These states are interpreted as
charged matter, and the root vectors $p$ are their weights with
respect to the unbroken gauge group.

A twisted string is one that closes only by imposing the space group
symmetry,
$$
Z^a(\tau, \sigma +2 \pi) = e^{(2 \pi i k v^a)} Z^a(\tau, \sigma) + n_{\alpha}e^{\alpha},
$$
i.e. by performing both twists and lattice shifts.  Thus, they must
be localized at the fixed points.  Each of these states is dependent
on the required number of twists, thus $k$ labels the $N-1$ twisted
sectors ($k=0$ corresponds to the untwisted sector).  Similarly, the
presence of Wilson lines is determined by the corresponding lattice
shifts required at each fixed point. Wilson lines affect the mass
equation for the left movers (as we will see below), so this has the
effect of changing the representations found at different fixed
points.  Modifying the boundary conditions for the twisted sector
changes the mode expansions for the right and left movers, which in
turn shifts the weights of the states, $q \rightarrow q+kv$ and $p
\rightarrow p + (k V + n_{\alpha} W_{\alpha})$.  As a result, the
level matching condition for the massless states now reads
$$
{1 \over 2}(q+kv)^2 - {1 \over 2} + \delta c = {1 \over 4} m^2_R = {1 \over 4} m^2_L = {1 \over 2} (p + kV + n_{\alpha} W_{\alpha})^2 + N_L -1 +\delta c = 0.
$$
$N_L$ is the number operator for the left movers, and is allowed to
be fractional as a consequence of a non-trivial twist.  To be more
specific, $N_L=\sum_a (\eta^a N_{La}+\bar{\eta}^a N_{La}^*)$, where
$\eta^a=kv^a$ mod 1 with $0 \leq \eta^a < 1$, $\bar{\eta}^a=-kv^a$
mod 1 with $0 \leq \bar{\eta}^a < 1$, and $N_{La}, N_{La}^*$ are
oscillator numbers of the left movers in the $z_a$ and $\bar{z}_a$
directions.  $\delta c$ is a shift in the zero point energy, and is
given by
$$
\delta c = {1 \over 2} \sum_{a=0}^3 \eta^a (1-\eta^a).
$$

Once the massless spectrum of the twisted sectors is calculated,
projection conditions must be applied.  Among the massless
representations, physical states are selected by the generalized GSO
projection operator.  In a theory with non-trivial Wilson lines, the
momentum shift is dependent on the fixed point under consideration
as discussed earlier.  Therefore, the GSO projection should be
applied to each state individually. This can be written:
$$
P(k,\gamma,n_{\alpha})={1 \over N} \sum^{N-1}_{l=0} [\Delta(k,\gamma,n_{\alpha})]^l
$$
with
$$
\Delta(k,\gamma,n_{\alpha})=\phi \gamma e^{2 \pi i[(P + kV+n_{\alpha}W_{\alpha}) \cdot (V+n_{\alpha}W_{\alpha})-(r+kv) \cdot v]}
$$
Here, $k$ labels the twisted sector, and the $n_{\alpha}$ label the
order of the Wilson lines relevant for the given state
(corresponding to the number of lattice shifts required for the
point to be invariant under the space group action).  $\gamma$ is
the eigenvalue of the state under the action of $k$ orbifold twists.
For prime orbifolds (e.g. $Z_3,Z_7$) this factor is trivial, $\gamma
=1$. For non-prime orbifolds, physical states are defined by linear
combinations of massless states living at fixed points which
transform into each other under the space group action.  These
physical states can be shown to have definite eigenvalue $\gamma =
e^{(2 \pi q_{\gamma})},q_{\gamma}=0,1/n,2/n...1$ under the rotation.
The oscillator phase is
$$
\phi=e^{2 \pi i \sum_a v_a (N_{La}-N_{La}^*)}.
$$
For any non-trivial phase $\Delta$, the contributions of $\Delta^l$
in the sum for $P$ will all add up to zero. Thus, only states
satisfying $\Delta(k, \gamma,n_{\alpha}) =1$ will survive the
projection.  Equivalently, the projection condition can be written:
$$
(P + kV+n_{\alpha}W_{\alpha}) \cdot (V+n_{\alpha}W_{\alpha})-(r+kv) \cdot v + \sum_a v_a (N_{La}-N_{La}^*) + q_{\gamma} = 0 \ {\rm mod} \ 1.
$$
For states with $q_{\gamma}=0$ (i.e. for prime orbifolds), one can
use the modular invariance equations to show that this condition is
in fact automatically satisfied for all states satisfying the mass
equation.  Thus, all massless representations of the prime orbifolds
are in fact physical states, and the GSO projector need not be
calculated.

The above construction provides the rules for calculating the entire
low-energy spectrum of the heterotic orbifold theory.  For
calculational convenience, we have automated the process using
Mathematica, and included it as an Appendix\footnote{The program
does not implement GSO projectors, and these must be checked by
hand.}. The required input is simply the orbifold twist vector, the
gauge shift, and the Wilson lines; the program will then check
modular invariance, calculate the gauge group and output the
surviving simple roots and Cartan matrix, and calculate the
surviving states in both the untwisted and twisted sectors,
displaying the highest weight representations in Dynkin label
notation.  The surviving gauge groups and representations may be
interpreted by comparison with, for example, the extensive tables of
\cite{slansky}.

\subsection{Search Strategy and Results}

Heterotic models based on $Z_N$ orbifolds are well known and have
been discussed extensively in the literature. There are a finite
number of gauge groups obtainable from $E_8 \times E_8$ for a
particular orbifold, and these have all been systematically
classified and their matter contents calculated.  In
\cite{Katsuki:1989cs}, the authors have tabulated the results for
every inequivalent modular invariant gauge shift (with no Wilson
lines) for each discrete orbifold.  Wilson lines complicate the
theory significantly, as they provide a mechanism to further break
down the gauge symmetries of the models as well as to change the
representations found at different fixed points, thereby greatly
increasing the number of inequivalent models.  Still, the rules are
well understood and a large number of these models have been
calculated. The prime orbifold $Z_3$ is particularly well known as
it has the simplest transformation properties under the orbifold
group.  Therefore, we have chosen the $Z_3$ orbifold as the starting
point for our search. Calculations of twisted sector states in $Z_3$
models are greatly simplified due to the fact that GSO projectors
need not be calculated if strong modular invariance is satisfied.
Conversely, the simplicity of the projectors in this case allow a
straightforward calculation of physical states, suggesting we may
employ weak modular invariance to ease the constraints on our
models.  We have elected to follow the latter approach.

There are only five possible breakings of $E_8$ by $N=3$ modular
invariant gauge shifts (without Wilson lines): $E_6 \times SU(3),\
SU(9),\ E_7 \times U(1),\ SO(14) \times U(1)$, and $E_8$ (unbroken).
Clearly, the $SU(5)$ factors of the Pentagon model would have to
arise from different $E_8$s, and there is only a limited number of
Wilson lines that would provide the desired symmetry.  However,
there is a very large number of ways to fit $SU(3)^4$ into $E_8
\times E_8$.  Furthermore, it would seem natural for the $Z_3$
symmetry of the trinification model to arise as the result of the
geometry of the orbifold.  Thus, we have elected to confine our
search to the particle spectrum of the Pyramid model.  That is, we
wish to find the low energy gauge group $SU(3)^4$ with matter
content
$$\begin{array}{ll}
3 \times (3,1,\bar{3};1)+(1,\bar{3},3;1)+(\bar{3},3,1;1) & \hspace{.2in} {\rm Standard \ Model \ Fermions} \nonumber \\
(3,1,1;\bar{3})+(\bar{3},1,1;3)& \hspace{.2in} \nonumber \\
(1,3,1;\bar{3})+(1,\bar{3},1;3)& \hspace{.2in} {\rm Trianons} \nonumber \\
(1,1,3;\bar{3})+(1,1,\bar{3};3)&   \hspace{.2in} \nonumber
\end{array}$$
on a $Z_3$ orbifold with twist vector $(1/3,1/3,-2/3)$.  The matter
content of standard trinification fits naturally into a $27$
representation of $E_6$. Thus, we will further assume that three of
the $SU(3)$ factors fit into an $E_6$ subgroup of a single $E_8$.
There is only one gauge shift that will break $E_8$ to $E_6 \times
SU(3)$ on a $Z_3$ orbifold, $V=(2/3,1/3,1/3,0,0,0,0,0)$ (there are
other modular invariant gauge shifts that have the same effect, but
they are all equivalent to the one listed by shifts in the lattice).
Of course, in our models $V$ has 16 components, 8 degrees of freedom
corresponding to each $E_8$.  The full vector must satisfy (strong)
modular invariance, and the condition $(V^2-v^2)=0$ mod 2 provides
little freedom for the last eight components.  Thus we will choose
$V=(2/3,1/3,1/3,0,0,0,0,0)(0,0,0,0,0,0,0,0)$, leaving the second
$E_8$ unbroken for the moment.  This will change with the addition
of Wilson lines.

In fact, realistic trinification models have been discovered under
these assumptions \cite{kimtrini}. These models do not,
unfortunately, exactly reproduce the spectrum of the Pyramid model.
In particular, while some of these do include a fourth $SU(3)$ gauge
group, none contain a full set of vector-like trianons.  However,
their models do provide useful guidelines for directing our search.
There are only two models listed in the tables of
\cite{Katsuki:1989cs} which give an $SU(3)^3$, but these do not have
the correct matter content.  Thus, one is forced to consider a model
with  Wilson lines.  In \cite{dynkinmethod} the authors have
classified all possible Wilson line breakings of $E_6 \times SU(3)$
on a $Z_3$ orbifold with one Wilson line, and have tabulated the
resulting gauge groups.  There are only two possibilities (up to
lattice shifts) for obtaining $SU(3)^3$.  They are
$$\begin{array}{ll}
W_1=(0,2/3,1/3,1/3,1/3,1/3,0,0)& \rightarrow  SU(3)^3 \times U(1)^2 \\
W_2=(5/3,1/3,1/3,1/3,1/3,1/3,0,0)& \rightarrow SU(3)^4
\end{array}$$
It would be convenient if the entire Pyramid model fit into a single
$E_8$, with the second $E_8$ remaining hidden.  For that to be true,
the gauge shift would break $E_8$ to the Pyramid $SU(3)$ times the
Standard Model GUT $E_6$, and Wilson lines (specifically $W_2$
above) would further break $E_6$ to the desired trinification
$SU(3)$.  Unfortunately, these choices do not produce the desired
spectrum, and it appears that the Pyramid $SU(3)$ will have to arise
from the second $E_8$. Phenomenologically this poses no problem; it
does however make obtaining this model quite difficult, due to the
fact that there are no fields charged under both $E_8$s in the
non-orbifolded heterotic string theory.  Since all chiral
representations obtained in the untwisted sector are merely a
subgroup of the entire $E_8 \times E_8$ adjoint which survive the
projection conditions, it is impossible to obtain representations
charged under both $E_8$s from the untwisted sector.  However,
because the momenta of the states existing at the fixed points of
the lattice are shifted by the presence of Wilson lines, it is
possible to obtain states charged under both $E_8$s in the twisted
sector.

Thus, our search strategy has been as follows.  We begin with the
$Z_3$ orbifold obtained from the gauge twist $(1/3,1/3,-2/3)$. We
wish to break the first $E_8$ to $E_6 \times SU(3)$ via the gauge
twist $V=(2/3,1/3,1/3,0,0,0,0,0)(\overrightarrow{0})$.  To obtain
standard trinification with no chiral exotics, we must further break
$E_6 \rightarrow SU(3)^3$ and $SU(3) \rightarrow U(1)^2$.  This can
be acheived by $W_1$ alone, or by a combination of $W_2$ and
additional Wilson lines, such as $W_3=(1/3,1/3,0,0,0,0,0,0)$.
Whichever we choose, we must also assign values for the eight
additional components corresponding to the second $E_8$ such that
the entire 16 component vector remains modular invariant.  The
second $E_8$ gauge group must be broken to $SU(3) \times G$, where
$G$ is some unspecified cofactor.  Phenomenologically the group $G$
is arbitrary as long as there are no fields charged simultaneously
under both it and the trinification group.

Because the lattice vector $e_1^i$ is equivalent to $e_2^i$ on each
two-torus $T^2_i$ by the action of the twist, the $Z_3$ orbifold can
sustain a maximum of three Wilson lines, one for each two-torus.  As
we have discussed, the presence of a Wilson line differentiates
states at different fixed points of the corresponding torus.
Because there are 27 fixed points on the $Z_3$ orbifold, the
multiplicity of twisted sector states in a model without Wilson
lines would be 27.  Models with one Wilson line will have twisted
sector multiplicity 9, two Wilson lines give multiplicity 3, and
three Wilson lines differentiates each fixed point individually.
Models with two Wilson lines seem to suggest a geometric explanation
for the family multiplicity of the Standard Model.  However, we are
constrained to find only a single generation of the trianons, and
are therefore led to consider models with three Wilson lines.  This
of course complicates the task of finding three trinification
generations.

Thus far we have fixed the gauge twist and have narrowed the
possibilities for one of the Wilson lines.  There still remains the
freedom to choose two additional Wilson Lines--each of which is a 16
dimensional vector.  Constraints are imposed due to the fact that
the Wilson lines must obey modular invariance and by the requirement
that we do not break the gauge symmetry of the first $E_8$ beyond
$SU(3)^3$.  Nevertheless, this still permits a vast number of models
to be calculated if we are to scroll through each possible vector in
succession (perhaps on the order of $>10^{10}$), making a
comprehensive search rather difficult.  At the present time we do
not have the computing power necessary to perform such a search,
though we would like to do so in the future.  Actually, it is
conceivable that the number of distinct Wilson lines is in fact much
smaller due to the fact that many will be equivalent up to lattice
shifts, but we have not found a way to use this fact to our
advantage at the current time.  Thus, to this point we have only
endeavored to follow the more modest approach of trial and error.

\begin{table}[!h!t!b]
\begin{center}
\begin{tabular}{|c|cc|}
\hline
Gauge Group& $SU(3)^3$ & $SU(3) \times SO(8)$\\
\hline
 $V$ &(2/3,1/3,1/3,0,0,0,0,0) & (0,0,0,0,0,0,0,0) \\
  $W_1$ &(0,2/3,1/3,1/3,1/3,1/3,0,0) &(2/3,0,0,0,0,0,0,0)  \\
 $W_2$ & (0,2/3,1/3,1/3,1/3,1/3,0,0) &(1/3,1/3,1/3,1/3,0,0,0,0) \\
 $W_3$ & (1/3,0,1/3,0,0,0,0,0) &(1/3,1/3,1/3,1/3,0,0,0,0) \\
\hline \hline
Spectrum & Particle & Vector Partner\\
\hline
 Trianons& $(\bar{3},1,1)(\bar{3},1)$ & \\
& $(1,3,1)(3,1)$& \\
 Trinification & $3 \times (1,3,\bar{3})(1,1)$ &  \\
& $3 \times (3,\bar{3},1)(1,1)$ &\\
& $3 \times (\bar{3},1,3)(1,1)$ & \\
 Higgs & $13 \times (3,1,1)(1,1)$&$ 7 \times (\bar{3},1,1)(1,1)$ \\
& $7 \times (1,3,1)(1,1)$ & $13 \times (1,\bar{3},1)(1,1) $\\
& $13 \times (1,1,3)(1,1)$& $13 \times (1,1,\bar{3})(1,1) $\\
 Exotics& $3 \times (1,3,3)(1,1)$  & \\
& $3 \times (\bar{3},1,\bar{3})(1,1)$ & \\
& $3 \times (3,\bar{3},1)(1,1)$ & \\
& $4 \times (1,1,1)(3,1)$ & \\
&$ 4 \times (1,1,1)(\bar{3},1)$ & \\
&$ 4 \times (1,1,1)(1,8)$ & \\
 Singlets& $ 17 \times (1,1,1)(1,1)$ & \\
\hline
\end{tabular}
\end{center}
\caption{$Z_3$ heterotic orbifold model 1.  Contains Standard Model trinification and chiral trianon-like particles}
\label{Tablehetero1}
\end{table}

\begin{table}[!h!t!b]
\begin{center}
\begin{tabular}{|c|cc|}
\hline
Gauge Group& $SU(3)^3$ & $SU(3) \times E_6$\\
\hline
 $V$ &(2/3,1/3,1/3,0,0,0,0,0) & (0,0,0,0,0,0,0,0) \\
  $W_1$ &(5/3,1/3,1/3,1/3,1/3,1/3,0,0) &(2/3,1/3,1/3,0,0,0,0,0)  \\
 $W_2$ & (2/3,1/3,1/3,0,0,0,0,0) &(2/3,1/3,1/3,0,0,0,0,0) \\
 $W_3$ & (1/3,1/3,0,0,0,0,0,0) &(2/3,1/3,1/3,0,0,0,0,0) \\
\hline \hline
Spectrum & Particle & Vector Partner\\
\hline
 Trianons& $(3,1,1)(3,1)$ & \\
& $(1,\bar{3},1)(\bar{3},1)$& \\
& $(1,1,\bar{3})(\bar{3},1)$ & \\
 Trinification & $ (1,3,\bar{3})(1,1)$ &  $ (1,\bar{3},3)(1,1)$ \\
& $(3,\bar{3},1)(1,1)$ & $ (\bar{3},3,1)(1,1)$\\
& $(\bar{3},1,3)(1,1) $ &  $ (3,1,\bar{3})(1,1)$\\
 Higgs & $7 \times (3,1,1)(1,1)$&$ 7 \times (\bar{3},1,1)(1,1)$ \\
& $7 \times (1,3,1)(1,1)$ & $7 \times (1,\bar{3},1)(1,1) $\\
& $7 \times (1,1,3)(1,1)$& $7 \times (1,1,\bar{3})(1,1) $\\
 Exotics&   $ (1,3,3)(1,1)$ &  $ (1,\bar{3},\bar{3})(1,1)$ \\
& $(3,3,1)(1,1)$ & $ (\bar{3},\bar{3},1)(1,1)$\\
& $(3,1,3)(1,1) $ &  $ (\bar{3},1,\bar{3})(1,1)$\\\
&$ 4 \times (1,1,1)(\bar{3},1)$ & \\
&$  (1,1,1)(1,27)$ & \\
&$(\bar{3},1,1)(3,1)$ & \\
 Singlets& $ 7 \times (1,1,1)(1,1)$ & \\
\hline
\end{tabular}
\end{center}
\caption{$Z_3$ heterotic orbifold model 2. Contains a {\it chiral} set of trianons and {\it vector}-like trinification.}
\label{Tablehetero2}
\end{table}

Unfortunately, we have not found anything resembling the complete
spectrum of the Pyramid model.  While a large number of models
contain the standard trinification spectrum (as we should expect
considering we have specifically chosen our gauge twist and Wilson
lines to enforce this), it is very difficult to obtain the trianons.
We believe this is due to the difficulty of finding particles
charged under both $E_8$s.  It is interesting to note that the
models we have found closest resembling the spectrum of the Pyramid
contain a chiral set of the trianon-like particles, but finding
their vector-like partners has proved elusive.  This is not entirely
surprising, considering that the heterotic orbifold models were
originally constructed to produce a chiral spectrum. This could even
be a general symptom of these models (and a failure for our
purposes), but a comprehensive search would have to be conducted to
know this for certain.

We present two interesting results in the tables \ref{Tablehetero1},
\ref{Tablehetero2}.  The first model is perhaps the most promising.
It contains three complete trinification generations, as well as a
number of Higgs-like fields. It also contains a single generation of
(an incomplete set of) chiral trianon-like particles, but it does
not contain their vector-like partners.  The model also contains a
few chiral exotics. The second model is interesting in that it
contains a single complete set of {\it chiral}-like trianons (i.e.
one half of the 6 total).  It also complains a completely {\it
vector}-like trinification spectrum.  It should be noted that the
GSO projectors have not been implemented on the spectra of these
models (the projectors will only project out states, and the spectra
are incomplete to begin with), and the spectra listed are therefore
anomalous.

The next step in our research will be to clearly establish the
number of inequivalent modular invariant Wilson lines for the $Z_3$
orbifold, and to perform a comprehensive search for the Pyramid
model spectrum.  If such a search fails to produce the desired
spectrum, we will be forced to perform a similar search in the other
$Z_N$ orbifolds, probably forcing us to abandon our desire for the $Z_3$ trinification
symmetry to be an artifact of the geometry of the manifold\footnote{It might still arise as a result of a non-prime orbifold, $Z_6=Z_2 \times Z_3$ or $Z_{12} = Z_3 \times Z_4$.}.
Regardless of the outcome, we are still interested in a future
search for the $SU(5) \times SU(5)$ gauge group and particle
spectrum of the Pentagon model.

\section{Concluding remarks}

Though we have not been able to rule out the existence of the
Pentagon model as a low energy effective field theory embedded in a
string theory, we have thus far had no success in constructing such
a model.  In each of the embedding structures we have explored, the
constraints imposed by our criterion have proven to be quite strict.
In part this is due to the size of the desired gauge groups, but we
believe that an even more restricting constraint is the requirement
that we find both chiral and vector-like particles in the spectrum.

This requirement posed a strict constraint on the geometry of the
two stacks of D6-branes supporting $SU(5)$ gauge groups in the type
IIA construction.  In fact, we found that no such structure was able
to satisfy the same equation for the complex structure moduli while
remaining consistent with RR-tadpole charge cancelation.  This
problem translates into a difficulty with maintaining supersymmetry.
While there may exist a more complicated geometry satisfying all of
our criteria, finding such a model proved to be beyond the scope of
our current search.  However, another approach we are currently
investigating is to embed the Pyramid model into an intersecting
D6-brane construction, though the results of this search are still
unclear.

We did discover a potential candidate for the existence of the
Pentagon model in the case of M-theory manifolds of $G_2$ holonomy,
though the proof of its existence is beyond our capabilities.
However, such a model does not support a pervasive $SU(5) \times
SU(5)$ symmetry throughout the $G_2$ manifold.  If we include this
criteria as a requirement for the model, we have shown that it
becomes quite difficult to obtain both the vector-like pentaquarks
and the chiral antisymmetric 10 representations of the GUT $SU(5)$.
This follows from the fact that both representations are found at a
singularity which resolves as $SO(20) \rightarrow SU(10) + o + 45 +
\bar{45}$.  We may break the SU(10) via Wilson lines, leaving us
with chiral particles in the representations $(5,5)+(1,10)+(10,1)$.
Therefore, in this construction, it is impossible to find
vector-like partners for the pentaquarks without simultaneously
producing vector-like partners for the 10s.

We have also shown that it is difficult to obtain the vector-like
trianons of the Pyramid model in a heterotic orbifold construction.
While we were able to find models with a standard trinification
spectrum, we found that the trianons must arise in the twisted
sector of a $Z_3$ orbifold due to the fact that they must come from
fields charged under both $E_8$s of the uncompactified heterotic
theory.  We did not perform a systematic search through all
modular invariant gauge shifts so we cannot make any conclusions
about the existence of the Pyramid spectrum in these models, but we
were unable to find the complete spectrum in our search and believe
it likely that no gauge shift in the $Z_3$ will give rise to a
vector-like set of trianons.  If this is indeed the case, we are
forced to abandon our hope that the $Z_3$ symmetry of the Pyramid
model arises as an artifact of the geometry.

Despite our limited success, there still remain many avenues in the
vast landscape of string models to explore.  We are especially
interested in continuing our search for the Pyramid model of TeV
physics in the contexts of each of the three string theories we have
investigated.  We also believe that models of intersecting branes in
type IIB theory and F-theory models might afford us the techniques
required to build our desired low energy effective theory.

\section{Acknowledgments}

I would like to thank my dissertation advisor T. Banks for extensive discussions about this work.  I would also like to thank H. Haber for discussions about group
theory and representations, and Ben Dundee for his help in
understanding heterotic orbifolds.  This research was supported in part by
DOE grant number DE-FG03-92ER40689.

%
%
%
%
%
%
\vspace{1in}


\end{document}